\documentclass[10pt,epsf]{article}

\usepackage{graphicx}
\usepackage{indentfirst}
\usepackage{cite}
\usepackage{color}

\begin{document}

\title{\bf{Thermodynamics of Three-dimensional Black Holes via Charged Particle Absorption}}

\date{}
\maketitle

\begin{center}
\author{Bogeun Gwak}$^a$\footnote{rasenis@sogang.ac.kr}, \author{Bum-Hoon Lee}$^{a, b, c}$\footnote{bhl@sogang.ac.kr}\\
\vskip 0.25in
$^{a}$\it{Center for Quantum Spacetime, Sogang University, Seoul 04107, Korea}\\
$^{b}$\it{Department of Physics, Sogang University, Seoul 04107, Korea}\\
$^{c}$\it{Asia Pacific Center for Theoretical Physics, Pohang 37673, Korea}\end{center}
\vskip 0.6in

{\abstract
{We have shown that changes occur in a (2+1)-dimensional charged black hole by adding a charged probe. The particle increases the entropy of the black hole and guarantees the second law of thermodynamics. The first law of thermodynamics is derived from the change in the black hole mass. Using the particle absorption, we test the extremal black hole and find out that the mass of the extremal black hole increases more than the electric charge. Therefore, the outer horizon of the black hole still exists. However, the extremal condition becomes non-extremal.}

\thispagestyle{empty}
\newpage
\setcounter{page}{1}
\section{Introduction}

The anti-de Sitter(AdS)/conformal field theory(CFT) correspondence\,\cite{Maldacena:1997re,Gubser:1998bc,Witten:1998qj,Aharony:1999ti} relates a gravity theory in a bulk AdS spacetime to a CFT on the AdS boundary. Since AdS black hole thermodynamics is observed in the dual CFT that resides on the boundary\,\cite{Witten:1998zw}, dual CFT properties can be investigated through those of AdS black holes. The instability of a Reissner-Nordstr\"{o}m-AdS (RN-AdS) black hole has been found in the presence of a charged scalar field below critical temperature\,\cite{Gubser:2008px}. In the AdS/CFT correspondence, the instability is interpreted for the CFT on the boundary as superconducting instability\,\cite{Hartnoll:2008vx,Hartnoll:2008kx}. Including superconductivity, the condensed matter theory (CMT) is described as a (1+1)-dimensional field theory; thus, the dual gravity theory in the bulk corresponds to the (2+1)-dimensional Ba\~{n}ados-Teitelboim-Zanelli (BTZ) black hole\,\cite{Banados:1992wn,Banados:1992gq}. In the AdS/CFT correspondence, the charged BTZ black hole \cite{Martinez:1999qi,Myung:2009sh} is important as a duality of the (1+1)-dimensional boundary CFT with a background electric charge\,\cite{Maity:2009zz,Chaturvedi:2013ova} and a holographic superconductor\,\cite{Ren:2010ha,Jensen:2010em,Andrade:2011sx,Chang:2014jna}. Through this duality, the dual CFT phases are similar to Fermi-Luttinger liquids\,\cite{Hung:2009qk,Davison:2013uha}.

Black hole entropy is the key to studying black hole thermodynamics which depends on the mass and momenta of the black hole. The addition of a particle can change the mass of a black hole\,\cite{Penrose:1971uk}, which is divided into reducible and irreducible masses\,\cite{ChrisRu}. The reducible mass corresponds to the rotational energy, which can be decreased\,\cite{Bardeen:1970zz,Penrose:1971uk}. The irreducible mass is the surface energy\,\cite{Smarr}, which always increases due to the conserved quantities of the particle. The square of irreducible mass is proportional to the surface area of the black hole\cite{BH1b}. The surface area is also interpreted as black hole entropy\,\cite{BH1b,BH1}. The microscopic origin of the black hole entropy has been explained through microstates\,\cite{York:1983zb,Zurek:1985gd,Frolov:1993ym,Larsen:1995ss} and has been obtained in the (2+1)-dimensional BTZ black hole. BTZ black hole entropy corresponds to the Cardy formula\,\cite{Cardy} for the asymptotic growth of states for the (1+1)-dimensional dual CFT\,\cite{BrHn1} that resides on the AdS$_3$ boundary\,\cite{Strom1}. Through the correspondence, changes in the BTZ black hole are interpreted in terms of the dual CFT microstates\,\cite{Gwak:2012hq}.

A black hole has a singularity veiled by its horizon. If the horizon is unstable, the singularity is easily exposed to our universe. However, this naked singularity violates cosmic censorship; thus, the instability of the black hole horizon is related to the violation of cosmic censorship. The stability of the black hole horizon has been investigated under particle absorption\,\cite{Wald:1974ge} where the Kerr black hole horizon is stable. The validity of cosmic censorship depends on the perturbation method. For example, the RN and Kerr-Newman black holes violate cosmic censorship\,\cite{Hubeny:1998ga,Jacobson:2009kt}, but cosmic censorship for the RN black hole is still valid in terms of backreaction\,\cite{Isoyama:2011ea}. This validity was recently investigated for Myers-Perry\,\cite{BouhmadiLopez:2010vc,Doukas:2010be,Gwak:2011rp} and AdS\,\cite{Gwak:2012hq,Rocha:2011wp,Rocha:2014gza,Rocha:2014jma,Gwak:2014xra,Gwak:2015fsa} black holes.

In this paper, we investigate the stability of the charged BTZ black hole horizon under particle absorption. The stability comes from satisfaction of the first law of thermodynamics and leads to the second law of thermodynamics. However, an extremal black hole becomes a non-extremal black hole; therefore, the mass of the extremal black hole increases larger than the electric charge of the black hole in this process.

The paper is organized as follows. In section~\ref{sec2}, we briefly review the charged BTZ black hole solution and thermodynamic properties. In section~\ref{sec3}, we show a change in black hole properties. By satisfying the first law of thermodynamics, the second law of thermodynamics can be obtained from the equation of motion of the particle. In section~\ref{sec4}, we test the extremal black hole by the particle. In section~\ref{sec5}, we summarize our results.

\section{The Charged BTZ Black Hole}\label{sec2}

A charged BTZ black hole is the solution of the (2+1)-dimensional Einstein gravity coupled with electromagnetism\,\cite{Martinez:1999qi}
\begin{eqnarray}
S=\int d^3 x \sqrt{-g}\left(\frac{R-2\Lambda}{2\kappa}-\frac{1}{4}F_{\mu\nu}F^{\mu\nu}\right)\,.
\end{eqnarray}
The gravitational constant $\kappa=8\pi G$ is set to be $\frac{1}{2}$, and the cosmological constant $\Lambda$ is related to the AdS radius $\ell$ by $-\frac{1}{\ell^2}$. In (2+1)-dimensional spacetime, the electric potential $A_\mu$ with a positive electric charge $Q$ are given
\begin{eqnarray}
A_{t}=Q\ln r\,.
\end{eqnarray}
The black hole metric has mass $M$ and electric charge $Q$
\begin{eqnarray}
\label{cbtzmetric}
ds^2=-f(r)dt^2+f^{-1} dr^2+r^2 d\phi^2\,,\,\,\,\,\, f(r)=-M+r^2-\frac{Q^2}{2} \ln r\,,
\end{eqnarray} 
with the dimensionless coordinates scaling on $\ell$. The electric potential $\Phi_H$ at the horizon are defined on the horizon $r_h$
\begin{eqnarray}
\Phi_h=\frac{\partial M}{\partial Q}{\Big |}_{r=r_h}=-Q\ln r_h\,.
\end{eqnarray}
The Hawking temperature and the Bekenstein-Hawking entropy are\cite{Banados:1992wn}
\begin{eqnarray}
T_H=\frac{1}{8\pi r_h}\left(4r_h^2-Q^2\right)\,,\,\,\,\,\,S_{BH}=4\pi r_h\,.
\end{eqnarray}

\section{Horizon Stability on a Particle}\label{sec3}

We perturb a black hole with a probe particle. The particle momenta $p_\mu$ infinitesimally change the black hole\,\cite{Wald:1974ge}. Since the momentum can be derived from the first-order geodesic equation of motion, we solve the equation of motion with the Hamilton-Jacobi method\,\cite{VPage1,VPage2,VPage3,VPage4}. The Hamiltonian $\mathcal{H}$ and the Hamilton-Jacobi action $S$ are
\begin{eqnarray}
\label{hjaction01}
\mathcal{H}=\frac{1}{2}g^{\mu\nu}(p_\mu-e\,A_\mu)(p_\nu-e\,A_\nu)\,,\,\,\,\,\, S=\frac{1}{2}m^2 \,\lambda - E\,t +L\,\phi+S_r(r)\,,
\end{eqnarray}
where the translation symmetries for $t$ and $\phi$ in Eq.~({\ref{cbtzmetric}}) give the conserved quantities $E$ and $L$. The affine parameter $\lambda$ has a conserved quantity $m$ as a particle mass. The Hamilton-Jacobi equation is given as
\begin{eqnarray}
-\frac{1}{2}m^2=\frac{1}{2}\left[-\frac{1}{f(r)}\left(-E-eQ\ln r\right)^2+f(r)\left(\frac{dS_r}{dr}\right)^2+\frac{L^2}{r^2}\right]\,,
\end{eqnarray}
where the momenta are defined as $p_\mu=\frac{\partial S}{\partial x^\mu}$\,. The Hamilton-Jacobi action in Eq.~(\ref{hjaction01}) is rewritten in terms of the redefined function $S_r(r)$
\begin{eqnarray}\label{radialEOM1}
&&S=\frac{1}{2}m^2 \,\lambda - E\,t +L\,\phi+\int{\sqrt{R(r)}\,dr}\,,\\ &&\sqrt{R(r)}\equiv\frac{dS_r}{dr}=\sqrt{\frac{1}{f(r)^2}\left(E+eQ\ln r\right)^2-\frac{1}{f(r)}\left(m^2+\frac{L^2}{r^2}\right)}\,.\nonumber
\end{eqnarray}
The particle radial momentum $p^r\equiv\dot{r}$ is described from the first-order radial equation obtained from Eq.~(\ref{radialEOM1}) for the given location and parameters
\begin{eqnarray}\label{radialEOM2}
(p^r)^2=\left(E+eQ\ln r\right)^2-f(r)\left(m^2+\frac{L^2}{r^2}\right)\,.
\end{eqnarray}
When a particle is absorbed in a charged BTZ black hole, the particle energy and charge merge with those of the black hole. The black hole undergoes infinitesimal changes in $\delta M$ and $\delta Q$. We treat the particle absorbed by the black hole when the particle touches the black hole horizon $r_h$. The corresponding black
hole mass and electric charge infinitesimally change by the particle
\begin{eqnarray}
\delta M = E\,,\quad\quad \delta Q =e\,.
\end{eqnarray}
The radial momentum is at the horizon
\begin{eqnarray}\label{1stb}
|p^r|=\delta M+Q\delta Q\ln r_h>0\,,
\end{eqnarray}
which is rewritten as the change on horizon $\delta r_h$
\begin{eqnarray}\label{chhori}
\delta r_h =\frac{2r_h}{4r_h^2-Q^2} \,|p^r|\,.
\end{eqnarray}
The BTZ black hole always satisfies $4r_h^2-Q^2\geq 0$, and the equality corresponds to the extremal case. Thus, the horizon radius increases and depends only on the particle radial momentum. The electric potential are
\begin{eqnarray}\label{angular1}
\delta \Phi_h=-\left(\ln r_h\right)\,e -\frac{2Q}{4r_h^2-Q^2}\,|p^r|\,.
\end{eqnarray}
The particle changes the black hole temperature
\begin{eqnarray}\label{tem1}
\delta T_H=\frac{1}{4\pi}\left[-\frac{Q\,e}{r_h}+\frac{2r_h}{4r_h^2- Q^2}\left(2+\frac{Q^2}{2r_h^2}\right)|p^r|\right],
\end{eqnarray}
where the black hole temperature depends on the particle momenta and charge. The infinitesimal change in the entropy depends only on the particle radial momentum
\begin{eqnarray}\label{ent03}
\delta S_{BH}=\frac{8\pi r_h}{4r_h^2-Q^2 }\left(\delta M+Q\delta Q\ln r\right)=\frac{8\pi r_h |p^r|}{4r_h^2-Q^2 }\geq 0\,,
\end{eqnarray} 
where the black hole entropy always increases when a particle is absorbed. This implies the second law of thermodynamics. When a particle is absorbed, the change in the horizon is always positive and independent of the electric charge of the particle; therefore, the black hole becomes larger than before. The first law of thermodynamics is obtained from Eq.~(\ref{1stb}) and (\ref{ent03})
\begin{eqnarray}
\delta M = \Phi_h \delta Q + T_H \delta S_{BH}\,.
\end{eqnarray}
Generally, the laws of thermodynamics are independently satisfied. However, the particle absorption guarantees the second law of thermodynamics, and it leads to satisfy the first law of thermodynamics.

\section{The Extremal Black Hole Case}\label{sec4}

The change in the extremal BTZ black hole is more drastic and has singular behavior. All properties are finite values, but the changes in response to the particle momenta are singular for the extremal condition. To understand this behaviors, we test the extremal black hole by the charged particle. The extremal black hole satisfies
\begin{eqnarray}
f(r_h)=0\,,\,\,\,\,\,\frac{df(r)}{dr}\biggl{|}_{r=r_h}=f'(r_h)=0\,,\,\,\,\,\,\frac{d^2f(r)}{dr^2}\biggl{|}_{r=r_h}=f''(r_h)>0\,,
\end{eqnarray}
where the horizon $r_h$ undergoes an infinitesimal increase $\delta r_h>0$ due to the particle momentum. The minimum value of the function $f(r)$ becomes negative at the minimum point $r_h+\delta r_e$
\begin{eqnarray}
f(r_h+\delta r_e)=-\delta M -Q\ln r_h \delta Q=-|p^r|<0\,.
\end{eqnarray}
Therefore, the particle cannot overcharge the extremal black hole. Absorbing the particle, the BTZ black hole becomes larger than before, and its horizon will be $r_h+\delta r_h$. The extremal condition changes into
\begin{eqnarray}
f(r_h+\delta r_h)\approx f(r_h)+f'(r_h)\delta r_h=0\,,\,\,\,\,\,f'(r_h+\delta r_h)\approx f'(r_h)+f''(r_h)\delta r_h>0\,,
\end{eqnarray}
where the changed horizon $r_h+\delta r_h$ is still the horizon, but it does not satisfy the extremal condition. Therefore, the extremality becomes non-extremal. The electric potential changes as $\delta\Phi_h<0$ in Eq.~(\ref{angular1}), because these are saturated to the largest values for given a black hole mass under the extremal condition. The change in the temperature is positive infinity as $\delta T_H > 0$ in Eq.~(\ref{tem1}). Thus, the temperature cannot be negative due to the particle, and the zero temperature of the extremal black hole acts as a kind of boundary for these singular behaviors. The extremal black hole entropy increases as shown in Eq.~(\ref{ent03}).

\section{Conclusion and Discussion}\label{sec5}

We tested the charged BTZ black hole thermodynamic properties by adding a charged probe. To describe the black hole properties as the probe momentum, the equations of motions were obtained through the Hamilton-Jacobi method. The equations of motion can be solved at the black hole horizon. The energy and electric charge of the particle can affect the corresponding black hole mass and charge. We solved the equations of motion at the black hole horizon. The change in the black hole mass is only depends on the radial momentum and electric charge of the particle. In this process, the entropy depends only on the particle radial momentum and always increases such as the second law of thermodynamics. The change in the black hole mass becomes the first law of thermodynamics using the change in the entropy.

We have tested whether the extremal black hole can be overcharged through particle absorption. Under the process, the mass of the black hole increases larger than the charge of the black hole, and the extremal black hole becomes non-extremal one. The black hole has two horizons and becomes larger than before. Therefore, the black hole horizon still exists. This behavior corresponds to the increase of the extremal black hole temperature.

\vskip 0.20in
{\bf Acknowledgments}
This research was supported by Basic Science Research Program through the National Research Foundation of Korea(NRF) funded by the Ministry of  Science, ICT \& Future Planning(2015R1C1A1A02037523) and the National Research Foundation of Korea(NRF) grant funded by the Korea government(MSIP)(No.~2014R1A2A1A010).


\begin{thebibliography}{99}

\bibitem{Maldacena:1997re} 
  J.~M.~Maldacena,
  Int.\ J.\ Theor.\ Phys.\  {\bf 38}, 1113 (1999)
  [Adv.\ Theor.\ Math.\ Phys.\  {\bf 2}, 231 (1998)].

\bibitem{Gubser:1998bc} 
  S.~S.~Gubser, I.~R.~Klebanov and A.~M.~Polyakov,
  Phys.\ Lett.\ B {\bf 428}, 105 (1998).

\bibitem{Witten:1998qj} 
  E.~Witten,
  Adv.\ Theor.\ Math.\ Phys.\  {\bf 2}, 253 (1998).

\bibitem{Aharony:1999ti} 
  O.~Aharony, S.~S.~Gubser, J.~M.~Maldacena, H.~Ooguri and Y.~Oz,
  Phys.\ Rept.\  {\bf 323}, 183 (2000).

\bibitem{Witten:1998zw} 
  E.~Witten,
  Adv.\ Theor.\ Math.\ Phys.\  {\bf 2}, 505 (1998).

\bibitem{Gubser:2008px} 
  S.~S.~Gubser,
  Phys.\ Rev.\ D {\bf 78}, 065034 (2008).

\bibitem{Hartnoll:2008vx} 
  S.~A.~Hartnoll, C.~P.~Herzog and G.~T.~Horowitz,
  Phys.\ Rev.\ Lett.\  {\bf 101}, 031601 (2008).

\bibitem{Hartnoll:2008kx} 
  S.~A.~Hartnoll, C.~P.~Herzog and G.~T.~Horowitz,
  JHEP {\bf 0812}, 015 (2008).

\bibitem{Banados:1992wn} 
  M.~Banados, C.~Teitelboim and J.~Zanelli,
  Phys.\ Rev.\ Lett.\  {\bf 69}, 1849 (1992).

\bibitem{Banados:1992gq} 
  M.~Banados, M.~Henneaux, C.~Teitelboim and J.~Zanelli,
  Phys.\ Rev.\ D {\bf 48}, 1506 (1993)
  [Erratum-ibid.\ D {\bf 88}, no. 6, 069902 (2013)].

\bibitem{Martinez:1999qi} 
  C.~Martinez, C.~Teitelboim and J.~Zanelli,
  Phys.\ Rev.\ D {\bf 61}, 104013 (2000).

\bibitem{Myung:2009sh} 
  Y.~S.~Myung, Y.~W.~Kim and Y.~J.~Park,
  Gen.\ Rel.\ Grav.\  {\bf 42}, 1919 (2010).

\bibitem{Maity:2009zz} 
  D.~Maity, S.~Sarkar, N.~Sircar, B.~Sathiapalan and R.~Shankar,
  Nucl.\ Phys.\ B {\bf 839}, 526 (2010).

\bibitem{Chaturvedi:2013ova} 
  P.~Chaturvedi and G.~Sengupta,
  Phys.\ Rev.\ D {\bf 90}, no. 4, 046002 (2014).

\bibitem{Ren:2010ha} 
  J.~Ren,
  JHEP {\bf 1011}, 055 (2010).

\bibitem{Jensen:2010em} 
  K.~Jensen,
  JHEP {\bf 1101}, 109 (2011).

\bibitem{Andrade:2011sx} 
  T.~Andrade, J.~I.~Jottar and R.~G.~Leigh,
  JHEP {\bf 1205}, 071 (2012).

\bibitem{Chang:2014jna} 
  H.~C.~Chang, M.~Fujita and M.~Kaminski,
  JHEP {\bf 1410}, 118 (2014);
  F.~M.~Haehl, R.~Loganayagam and M.~Rangamani,
  JHEP {\bf 1403}, 034 (2014).

\bibitem{Hung:2009qk} 
  L.~Y.~Hung and A.~Sinha,
  JHEP {\bf 1001}, 114 (2010);
  V.~Balasubramanian, I.~Garcia-Etxebarria, F.~Larsen and J.~Simon,
  Phys.\ Rev.\ D {\bf 84}, 126012 (2011).

\bibitem{Davison:2013uha} 
  R.~A.~Davison, M.~Goykhman and A.~Parnachev,
  JHEP {\bf 1407}, 109 (2014).



\bibitem{Penrose:1971uk} 
  R.~Penrose and R.~M.~Floyd,
  Nature {\bf 229}, 177 (1971).

\bibitem{ChrisRu}
  D.~Christodoulou, 
  Phys.\ Rev.\ Lett.\  {\bf 25}, 1596 (1970);
  D.~Christodoulou and R.~Ruffini, 
  Phys.\ Rev.\  D {\bf 4}, 3552 (1971).

\bibitem{Bardeen:1970zz} 
  J.~M.~Bardeen,
  Nature {\bf 226}, 64 (1970).

\bibitem{Smarr}
  L.~Smarr, 
  Phys.\ Rev.\ Lett.\  {\bf 30}, 71 (1973).

\bibitem{BH1b}
  J.~D.~Bekenstein, 
  Phys.\ Rev.\  D {\bf 7}, 2333 (1973);
  J.~D.~Bekenstein, 
  Phys.\ Rev.\  D {\bf 9}, 3292 (1974);
  S.~W.~Hawking, 
  Commun.\ Math.\ Phys.\  {\bf 43}, 199 (1975).

\bibitem{BH1}
  J.~D.~Bekenstein, 
  Phys.\ Rev.\  D {\bf 9}, 3292 (1974);
  S.~W.~Hawking, 
  Commun.\ Math.\ Phys.\  {\bf 43}, 199 (1975).

\bibitem{York:1983zb} 
  J.~W.~York, Jr.,
  Phys.\ Rev.\ D {\bf 28}, 2929 (1983).

\bibitem{Zurek:1985gd} 
  W.~H.~Zurek and K.~S.~Thorne,
  Phys.\ Rev.\ Lett.\  {\bf 54}, 2171 (1985).

\bibitem{Frolov:1993ym} 
  V.~P.~Frolov and I.~Novikov,
  Phys.\ Rev.\ D {\bf 48}, 4545 (1993).

\bibitem{Larsen:1995ss} 
  F.~Larsen and F.~Wilczek,
  Phys.\ Lett.\ B {\bf 375}, 37 (1996).

\bibitem{Cardy}
  J.~L.~Cardy, 
  Nucl.\ Phys.\  B {\bf 270}, 186 (1986).

\bibitem{BrHn1}
  J.~D.~Brown and M.~Henneaux, 
  Commun.\ Math.\ Phys.\  {\bf 104}, 207 (1986);
  O.~Coussaert and M.~Henneaux, 
  Phys.\ Rev.\ Lett.\  {\bf 72}, 183 (1994).

\bibitem{Strom1}
  A.~Strominger, 
  JHEP {\bf 9802}, 009 (1998).

\bibitem{Gwak:2012hq} 
  B.~Gwak and B.~-H.~Lee,
  Class.\ Quant.\ Grav.\  {\bf 29}, 175011 (2012).



\bibitem{Wald:1974ge} 
  R.~M.~Wald,
  Annals Phys.\  {\bf 82}, 548 (1974).






\bibitem{Hubeny:1998ga} 
  V.~E.~Hubeny,
  Phys.\ Rev.\ D {\bf 59}, 064013 (1999).

\bibitem{Jacobson:2009kt} 
  T.~Jacobson and T.~P.~Sotiriou,
  Phys.\ Rev.\ Lett.\  {\bf 103}, 141101 (2009)
  [Erratum-ibid.\  {\bf 103}, 209903 (2009)];
  A.~Saa and R.~Santarelli,
  Phys.\ Rev.\ D {\bf 84}, 027501 (2011);
  S.~Gao and Y.~Zhang,
  Phys.\ Rev.\ D {\bf 87}, no. 4, 044028 (2013).

\bibitem{Isoyama:2011ea} 
  S.~Isoyama, N.~Sago and T.~Tanaka,
  Phys.\ Rev.\ D {\bf 84}, 124024 (2011).

\bibitem{BouhmadiLopez:2010vc} 
  M.~Bouhmadi-Lopez, V.~Cardoso, A.~Nerozzi and J.~V.~Rocha,
  Phys.\ Rev.\ D {\bf 81}, 084051 (2010).

\bibitem{Doukas:2010be} 
  J.~Doukas,
  Phys.\ Rev.\ D {\bf 84}, 064046 (2011).

\bibitem{Gwak:2011rp} 
  B.~Gwak and B.~-H.~Lee,
  Phys.\ Rev.\ D {\bf 84}, 084049 (2011).

\bibitem{Rocha:2011wp} 
  J.~V.~Rocha and V.~Cardoso,
  Phys.\ Rev.\ D {\bf 83}, 104037 (2011).

\bibitem{Rocha:2014gza} 
  J.~V.~Rocha, R.~Santarelli and T.~Delsate,
  Phys.\ Rev.\ D {\bf 89}, no. 10, 104006 (2014).

\bibitem{Rocha:2014jma} 
  J.~V.~Rocha and R.~Santarelli,
  Phys.\ Rev.\ D {\bf 89}, no. 6, 064065 (2014).

\bibitem{Gwak:2014xra} 
  B.~Gwak and B.~-H.~Lee,
  Phys.\ Rev.\ D {\bf 91}, no. 6, 064020 (2015).

\bibitem{Gwak:2015fsa} 
  B.~Gwak and B.~-H.~Lee,
  arXiv:1509.06691.

\bibitem{VPage1}
  M.~Vasudevan, K.~A.~Stevens and D.~N.~Page,
  Class.\ Quant.\ Grav.\  {\bf 22}, 1469 (2005).

\bibitem{VPage2}
  D.~N.~Page, D.~Kubiznak, M.~Vasudevan and P.~Krtous,
  Phys.\ Rev.\ Lett.\  {\bf 98}, 061102 (2007).

\bibitem{VPage3}
  V.~P.~Frolov, P.~Krtous and D.~Kubiznak,
  JHEP {\bf 0702}, 005 (2007).

\bibitem{VPage4}
  M.~Vasudevan, K.~A.~Stevens and D.~N.~Page,
  Class.\ Quant.\ Grav.\  {\bf 22}, 339 (2005);
  M.~Vasudevan and K.~A.~Stevens,
  Phys.\ Rev.\  D {\bf 72}, 124008 (2005).

\end{thebibliography}
\end{document}